\documentclass[conference,compsoc]{IEEEtran}
\ifCLASSOPTIONcompsoc
  \usepackage[nocompress]{cite}
\else
  \usepackage{cite}
\fi
\ifCLASSINFOpdf
\else
\fi

\usepackage[]{subcaption}
\usepackage{caption}
\usepackage{color}
\usepackage{xcolor}
\usepackage{xspace}
\usepackage{comment}
\usepackage{hyperref}
\usepackage{array}
\usepackage{multirow}
\hypersetup{
    colorlinks=true,
    linkcolor=red,
    filecolor=red,      
    urlcolor=red,
    citecolor=red
    }
\usepackage{longtable}
\usepackage{tabularx}
\usepackage{url}
\usepackage[flushleft]{threeparttable}
\usepackage[online]{threeparttablex}
\usepackage{enumitem}
\setlist{itemsep=0pt,parsep=0pt}
\usepackage[export]{adjustbox}
\usepackage[english]{babel}
\usepackage{xcolor} 
\usepackage[most]{tcolorbox} 
\usepackage{soul}
\usepackage{lipsum}

\newcommand{\claudeAccuracy}{\textsf{$93.8\%$}\xspace}
\newcommand{\claudePrecision}{\textsf{$92\%$}\xspace}
\newcommand{\claudeRecall}{\textsf{$94\%$}\xspace}
\newcommand{\claudeSpecificity}{\textsf{$93\%$}\xspace}

\begin{document}
%
\title{ThumbnailTruth: A Multi-Modal LLM Approach for Detecting Misleading YouTube Thumbnails Across Diverse Cultural Settings}

\author{
    {\large Wajiha Naveed, Zartash Afzal Uzmi, Zafar Ayyub Qazi} \\
    {\small Department of Computer Science} \\
    {\small Lahore University of Management Sciences} \\
    {\small Pakistan}
}


%


\maketitle

\begin{abstract}
Misleading video thumbnails on platforms like YouTube are a pervasive problem, undermining user trust and platform integrity. This paper proposes a novel multi-modal detection pipeline that uses Large Language Models (LLMs) to flag misleading thumbnails. We first construct a comprehensive dataset of 2,843 videos from eight countries, including 1,359 misleading thumbnail videos that collectively amassed over 7.6 billion views---providing a unique cross-cultural perspective on this global issue.
Our detection pipeline integrates video-to-text descriptions, thumbnail images, and subtitle transcripts to holistically analyze content and flag misleading thumbnails. Through extensive experimentation and prompt engineering, we evaluate the performance of state-of-the-art LLMs, including GPT-4o, GPT-4o Mini, Claude 3.5 Sonnet, and Gemini-1.5 Flash. Our findings show the effectiveness of LLMs in identifying misleading thumbnails, with Claude 3.5 Sonnet consistently showing strong performance, achieving an accuracy of 93.8\%, precision over 92\%, and recall exceeding 94\% in certain scenarios.
We discuss the implications of our findings for content moderation, user experience, and the ethical considerations of deploying such systems at scale. Our findings pave the way for more transparent, trustworthy video platforms and stronger content integrity for audiences worldwide.

\end{abstract}


%
\IEEEpeerreviewmaketitle

\section{INTRODUCTION}
\label{sec:intro}
In today's digital landscape, user-generated video content has become a dominant form of communication and entertainment. Platforms like YouTube, with over 2 billion logged-in monthly users, serve as global hubs for information sharing and creative expression. However, this vast ecosystem faces a persistent challenge: the proliferation of misleading thumbnail images. These deceptive previews, designed to entice clicks rather than accurately represent content, not only undermine user trust but also contribute to the spread of misinformation and clickbait culture.

The detection of misleading thumbnail videos is a critical problem with far-reaching implications for online platforms and their users. According to a recent study~\cite{srinivasan2021impact}, misleading thumbnails can lead to a 14\% increase in click-through rates compared to accurate thumbnails, incentivizing content creators to employ deceptive practices. Furthermore, a survey conducted by the Pew Research Center found that 64\% of adults have encountered misleading content online, with video thumbnails being a significant contributor to this issue~\cite{pew2022misinformation}. The prevalence of such content not only erodes user trust but also impacts the spread of misinformation and users perception of reality.

Detecting misleading thumbnails presents a multifaceted challenge for content platforms. The sheer volume of content uploaded daily—with over 500 hours of video uploaded to YouTube every minute—makes manual review impractical~\cite{youtube2024press}. Moreover, the subjective nature of what constitutes a misleading thumbnail, coupled with cultural and contextual nuances, further complicates automated detection efforts. Traditional image recognition techniques often fall short in capturing the nuanced relationship between a thumbnail and its corresponding video content. As a result, platforms struggle to effectively identify and moderate misleading thumbnails at scale, relying heavily on user reports and limited automated systems that may not capture the full spectrum of deceptive practices.

Large Language Models (LLMs) emerge as a promising solution to this complex problem. Their ability to understand and process multimodal inputs—combining text, images, and contextual information—makes them well-suited for the task of misleading thumbnail detection. LLMs can analyze the semantic relationship between a thumbnail, video title, and actual content, potentially identifying discrepancies that traditional methods might miss. Furthermore, their capacity for few-shot learning and adaptability to diverse contexts could address the challenge of detecting misleading thumbnails across different cultures and content types. As LLMs continue to advance in their multimodal capabilities, they offer a scalable and potentially more accurate approach to tackling the pervasive issue of misleading thumbnails in online video platforms.

This paper addresses this critical problem by developing and evaluating a novel approach to detect misleading thumbnails using state-of-the-art LLMs. Our study makes several key contributions:

\begin{itemize}[leftmargin=*,topsep=0pt]
\setlength{\itemsep}{0pt}
\setlength{\parskip}{0pt} 
    
\item \noindent \textbf{Comprehensive Dataset:} We have compiled a diverse and substantial dataset comprising 2,843 videos from YouTube across eight countries, evenly split between misleading and non-misleading thumbnails videos. Specifically, our dataset includes 1,359 misleading thumbnail videos, which collectively garnered over 7.6 billion views. Notably, each of the top 10 misleading thumbnail videos in our dataset had amassed over 100 million views individually. This extensive reach underscores the widespread impact of misleading thumbnails and the urgent need for effective detection methods. Our dataset provides a robust foundation for training and evaluating our detection models across various cultural contexts and content types. We have made the dataset, annotation codebook, scripts and a subset of misleading thumbnails publicly available via our GitHub repository for further replication and understanding:
\url{https://github.com/wajihanaveed/ThumbnailTruth.git}.
\item \noindent \textbf{Multi-Modal Analysis:} Our approach integrates video-to-text descriptions, thumbnail images, and subtitle information, offering a holistic analysis of content discrepancies. This multi-modal strategy enables a more nuanced and accurate assessment of whether a thumbnail is deceptive.

\item \noindent \textbf{Multiple LLMs Evaluation:} We conduct a thorough evaluation of four state-of-the-art LLMs: GPT-4o, GPT-4o Mini, Claude 3.5 Sonnet, and Gemini-1.5 Flash. This comparative analysis provides insights into the relative strengths and capabilities of these models in tackling the complex task of misleading thumbnail detection.

\item \noindent \textbf{Improvements from Prompt Engineering}: Our study explores the efficacy of various prompt engineering techniques, including chain-of-thought reasoning, fixed few-shot examples, and RAG-based dynamic few-shot examples. Through careful ablation studies, we quantify the performance benefits of each prompting strategy.

\item \noindent \textbf{Comparison with Task-Specific Solutions:}  
Our best performing LLM-based configuration---Claude 3.5 Sonnet with dynamic few-shot prompts—was benchmarked against \textsc{Checker} \cite{xie2021checker}, the leading supervised multimodal pipeline for detecting misleading thumbnails. Claude matched or outperformed \textsc{Checker} on every metric, showing that prompt-based LLMs can reach state-of-the-art accuracy without task-specific model training, offering a more flexible, easily deployable alternative for content-moderation workloads

\end{itemize}

We comprehensively evaluated each model's performance using four complementary metrics: accuracy, precision, recall, and specificity, providing a balanced view of their respective strengths. Our results confirm the promise of LLM-based detection. Among the models tested, Claude 3.5 Sonnet demonstrated superior performance across all metrics and prompt settings. Specifically, our observations include:
\begin{itemize}[leftmargin=*,topsep=0pt]
\setlength{\itemsep}{0pt}
\setlength{\parskip}{0pt} 
    
\item \textbf{Robust accuracy:} With dynamic few-shot prompts, Claude 3.5 Sonnet reaches \claudeAccuracy accuracy. For content moderators, this means most thumbnails are triaged correctly, sharply reducing manual review volume.
\item \textbf{High precision:} Claude 3.5 Sonnet maintains precision above \claudePrecision across all prompt styles, so the vast majority of flagged thumbnails are truly deceptive. This keeps false alarms low and sustains reviewer trust.
\item \textbf{High recall:} GPT-4o Mini and Claude 3.5 Sonnet exceed \claudeRecall in several settings, capturing most misleading thumbnails and preventing harmful content from slipping through automated checks.
\item \textbf{Strong specificity:} Claude 3.5 Sonnet consistently delivers specificity above \claudeSpecificity, rarely mislabeling legitimate thumbnails. This protects honest creators from unwarranted takedowns and lets moderators focus on genuine violations.
\item \textbf{Cultural variation:} Accuracy varies by country, underscoring the need for culturally adaptive prompts (e.g., dynamic examples in local languages). 
\end{itemize}
Our results indicate that LLMs hold promise for detecting misleading thumbnails, potentially strengthening platform integrity and user trust. Effective deployment will hinge on tuning false-positive thresholds, providing transparent appeals, and continually adapting to evolving manipulations—challenges we discuss in \S\ref{sec:discussion}.

The remainder of the paper covers our methodology, results with ablations and model benchmarks, and the implications for real-world deployment, content moderation, and user experience.


\section{METHODOLOGY}

This section outlines the systematic approach used to investigate misleading thumbnails across multiple countries. We describe key phases, including dataset construction, data processing, and analysis, which form the basis for evaluating LLMs in detecting misleading thumbnails. We detail the selection of countries, YouTube video collection, thumbnail extraction, subtitle retrieval, video downloading, and descriptive text generation. These steps contribute to a comprehensive dataset for analysis, ensuring a robust and replicable evaluation of LLM performance in detecting misleading thumbnails.

\subsection{Country Selection}
To ensure a broad representation of content and cultural practices related to misleading thumbnails, we sampled videos from eight countries—four developing and four developed—drawn from the 20 nations with the largest YouTube audiences. Countries were classified by real-GDP growth using the UN World Economic Situation and Prospects 2024 report \cite{statista2024youtubeusers}.
\begin{itemize}[leftmargin=*,topsep=0pt]
\setlength{\itemsep}{0pt}
\item \textbf{Developing:} Brazil, Pakistan, Indonesia, Mexico
\item \textbf{Developed:} USA, UK, Spain, Italy
\end{itemize}
This balanced sample supports meaningful comparisons between high-income and middle-income economies when analyzing misleading thumbnails.




\subsection{Dataset Construction}
To compile a comprehensive collection of Misleading Thumbnail Videos (MTVs), we employed a multi-step approach. First, we utilized Virtual Private Networks (VPNs) to simulate each country’s location, ensuring that our searches were region-specific. To maintain consistency and reduce the impact of personalized search results, all searches were conducted using Google Chrome's incognito mode.

Our primary search strategy was guided by Google Trends, using popular search terms relevant to each country. In Pakistan, we noticed that the letter ``f" in trending searches led to the discovery of several MTVs. Additionally, during our search experiments, we accidentally found that using a period (“.”) also uncovered many MTVs. These insights prompted us to incorporate random character searches (e.g., “f” and “.”) into our methodology, which, although yielding fewer results, introduced a unique dimension to our dataset.

Videos were collected from both the main search results page and the recommendation panel that appears when watching a video. Prior research suggests that viewing an MTV triggers YouTube's algorithm to recommend more MTVs, which helped expand our collection \cite{hussein2020measuring}. To ensure balanced analysis, we also compiled a dataset of NMTVs for comparative evaluation. For videos featuring non-English text—whether in thumbnails or subtitles—we used Google Translate for consistent interpretation across modalities.

Our initial dataset consisted of 3,200 videos, with 200 MTVS and 200 NMTVs sourced from each of eight countries. The annotation was performed by two individuals: one of the authors who curated the dataset and a trained university graduate. Both annotators followed a written codebook, with a detailed set of guidelines. The codebook defined misleading thumbnails as those exhibiting exaggeration, false promises, or thematic mismatch with the video content, while non-misleading thumbnails accurately reflected the video's main topic.

To reduce subjectivity, the codebook also included instructions for handling borderline cases. Thumbnails containing minor exaggeration but maintaining thematic alignment were labeled as non-misleading, whereas thumbnails that misrepresented the video's central theme were labeled as misleading. Annotators reviewed both the thumbnail and the corresponding video before assigning labels. We measured inter-annotator agreement using Cohen’s Kappa, which yielded a score of 0.9633, indicating near-perfect agreement. Only videos where both annotators agreed were retained. After removing cases with label disagreements, unavailable videos, or technical issues during video processing (e.g. video not downloading), the final dataset consisted of 2,843 videos, including 1,359 MTVs and 1,484 NMTVs.

\subsection{Data Processing}

To prepare our dataset for evaluation, we extracted three primary modalities from each video: the thumbnail image, subtitles, and a generated video-to-text description. These inputs were selected to reflect the information a typical viewer is most likely to encounter prior to or during early video engagement.

We excluded indirect social signals—such as comments, likes, and view metrics—for two key reasons. First, our objective is to ideally enable pre-hoc moderation, identifying misleading thumbnails before the video is uploaded and made publicly accessible—at which point engagement signals are not yet available. Relying on such post-hoc indicators would require user exposure, undermining the preventive intent of our approach. Second, prior research shows that users often choose not to engage with misleading or deceptive content, leading to sparse and unreliable feedback. For example, adolescents frequently scroll past questionable videos without commenting or reacting \cite{lindstol2023adolescents}. 

We also exclude video titles and descriptions. Although titles sometimes contain click-bait phrasing, prior work indicates they usually align with the video itself. In Qu et al. \cite{qu2018youtube} small-scale annotation study, nine of 87 ostensibly non-clickbait videos had “click-bait-looking” titles, yet none were actually misleading once the video was watched. In our own dataset, truly deceptive titles were similarly scarce, while descriptions were often empty, boiler-plate, or purely promotional—offering little semantic signal. In contrast, thumbnails, subtitles, and automatically generated video-to-text summaries, expose the imagery and narrative that viewers encounter before clicking “play.” These modalities therefore provide richer and more interpretable cues for detecting thumbnail–content mismatches, making them the focus of our multimodal detection pipeline.

\noindent \textbf{Thumbnail Extraction.} Thumbnails were downloaded using the standard YouTube URL format: \textit{https://img.youtube.com/vi/\{video\_id\}\/hqdefault.jpg}, retrieving the default high-quality thumbnail. These images were stored on Google Cloud Platform for evaluation with the Gemini model and locally for evaluations with Claude, GPT-4o, and GPT-4o-mini.

\noindent \textbf{Subtitle Retrieval.} A Python script was developed to automate the retrieval of YouTube video transcripts using the YouTube Data API \cite{google2024youtubeapi}. When subtitles were available in non-English languages, we translated them into English using Google Translate. Videos without auto-generated subtitles were retained in the dataset to maintain diversity and consistency.

\noindent \textbf{Video Download.} Videos were downloaded using the pytubefix library \cite{juanbindez2024pytubefix}. For videos exceeding 30 minutes, we limited the analysis to the first 29 minutes and 55 seconds for both subtitle extraction and video-to-text generation. This restriction ensures uniformity in our analysis and aligns with the processing constraints of tools such as Twelve Labs (30-minute limit) and Gemini 1.5 Flash (50-minute limit). To optimize storage, all videos were downloaded at a resolution of 360p and uploaded to Google Cloud Platform, Twelve Labs, and stored locally for further processing.

\noindent \textbf{Video Description Generation.}  
Video-to-text descriptions—referred to as video descriptions—were generated using Gemini \cite{google2024geminiflash}, Claude \cite{google2024claude35sonnet}, and Twelve Labs \cite{twelvelabs2024}, leveraging their advanced semantic understanding capabilities. These descriptions served as structured textual representations of the video content, enabling models to reason over scenes in conjunction with subtitles.

Rather than inputting raw video content into the classification prompt, we opted to generate concise video descriptions instead. This decision was driven by our observation that model accuracy declined with increased token length. Our findings align with recent work by Databricks \cite{databricks2024longrag}, which highlights performance degradation in LLMs as context length increases. Peng et al. \cite{hsieh2024rulerwhatsrealcontext} further quantify that usable context is often far shorter than the claimed maximum, reinforcing the need for brevity. Video descriptions thus offered a concise yet nuanced alternative to full-length video inputs, balancing informativeness with token efficiency.

After experimenting with various prompts, we found the following to be the most effective for Gemini and Twelve Labs, which support full video input:

\textit{“Watch the video and provide a detailed description. Break down the content scene by scene, focusing on key actions, visuals, and emotions.”}

These prompts were designed to elicit contextually rich, temporally grounded descriptions. By encouraging models to reason through scene transitions and emotional cues, the outputs captured not just surface-level content but the underlying narrative flow—critical for interpreting thumbnails whose meaning depends on broader context.

Since Claude 3.5 Sonnet does not support direct video input and limits inputs to 20 images, we extracted 20 evenly spaced frames and supplied them as input. The following prompt was used with Claude to synthesize a narrative description:

\textit{“Consider these frames as continuous scenes from a video. Provide a detailed description of the video content, breaking it down scene by scene. Focus on key actions, visuals, emotions, and any notable details. Describe it as if you are watching the full video, ensuring that the narrative is cohesive and captures the flow of the scenes.”}

By explicitly instructing the model to “consider these frames as continuous scenes from a video” and to “describe it as if you are watching the full video,” we encouraged Claude to interpret the stills as temporally linked and generate a coherent, scene-by-scene description. Without this framing, Claude tended to treat each frame in isolation—using phrases like “the frames show...” rather than narrating holistically. In contrast, models with video input naturally responded with “the video shows...”. By preserving the core of the prompt and adding temporal framing, Claude’s outputs closely mirrored those of video-aware models in both coherence and narrative structure.

Including video descriptions alongside subtitles addressed contextual gaps, especially in cases where subtitles were sparse or missing. Descriptions provided alternative context and, when subtitles were present, complemented them with a more cohesive summary. For example, in a video demonstrating how to create a Cristiano Ronaldo poster using Photoshop, sparse subtitles led the models to misclassify the thumbnail—featuring Ronaldo—as misleading. The generated description clarified the context, enabling correct classification.

Though descriptions occasionally introduced minor inaccuracies, they consistently improved contextual understanding. Both preliminary testing and our ablation study showed higher misclassification rates when descriptions were omitted, especially for ambiguous cases. Overall, this multimodal pipeline—integrating thumbnails, subtitles, and generated descriptions—enabled the construction of a robust dataset for evaluating LLM-based detection of misleading thumbnails.

\subsection{Prompts}
We explored various prompting strategies, as past research indicates that the structure and design of prompts significantly influence the reasoning performance of LLMs \cite{promptingguide2024techniques, kojima2023largelanguagemodelszeroshot}. Our experiments included three types of prompts: a Zero-shot prompt, followed by refinements into Fixed Few-shot and Dynamic Few-shot prompts. All three prompting strategies followed a clearly defined set of steps for classifying YouTube thumbnails. These instructions guided the LLMs in comparing the thumbnails with the actual video content and determining whether the thumbnails were misleading.

\noindent \textbf{Zero-Shot Prompt.}  
In the zero-shot setting, where the model received no prior examples, we instructed the LLM to analyze the relationship between a YouTube video’s thumbnail and its actual content using a single input instance at a time. To support the reasoning process, the model was guided by a structured prompt containing explicit step-by-step instructions. This mirrors the principles of zero-shot chain-of-thought prompting, where structured reasoning can emerge even without training examples, as demonstrated by Kojima et al. \cite{kojima2023largelanguagemodelszeroshot}. To ensure consistency in interpretation, we explicitly defined what constitutes a ``misleading thumbnail" within the prompt. Our definition drew on multiple sources. We referenced YouTube’s own policy, which describes misleading thumbnails as those that ``mislead viewers to think they’re about to view something that’s not in the video.” We further incorporated insights from prior research, which characterize clickbait thumbnails as those that “deviate substantially from [the] content” they represent \cite{zannettou2018clickbait}.These sources, along with our structured annotation guidelines, informed the criteria presented to the model and shaped the examples used in few-shot settings.
For each video, the model examined the thumbnail, reviewed the subtitles (if available) and the video description, and then assessed whether the thumbnail accurately reflected the content or relied on tactics such as exaggeration or clickbait.

Below is the exact prompt template used:
\begin{quote}
\textbf{Task:} Analyze the provided information about a YouTube video and determine whether its thumbnail is misleading or not misleading. You will be given the following information:
\begin{itemize}
\item The video's thumbnail
\item The video's subtitles
\item A text description of the video content
\end{itemize}

\textbf{Steps to Follow:}
\begin{enumerate}
\item Carefully examine the thumbnail.
\item Read through the video subtitles and content description.
\item Compare the thumbnail to the actual video content.
\item Determine if the thumbnail accurately represents the video's main topic or content.
\item Assess whether the thumbnail uses exaggeration, false promises, or clickbait tactics.
\item Categorize the thumbnail as either \textit{``Misleading"} or \textit{``Not Misleading"}.
\item Provide a brief explanation for your decision.
\end{enumerate}
Analyze the provided information and categorize the thumbnail as ``Misleading" or ``Not Misleading", followed by a brief explanation for your decision.

\textbf{Inputs:} \\
Video Description: \{\texttt{video\_description}\} \\
Video Subtitles: \{\texttt{video\_subtitles}\}
\end{quote}
This structured prompt enabled a controlled evaluation of the model’s performance without external guidance, serving as a baseline for subsequent prompting strategies.

\noindent \textbf{Fixed Few-Shot Prompt.}  
To improve classification consistency and reduce ambiguity, we introduced a fixed few-shot version of the same task. This prompt retained the structure and instructions of the zero-shot version but appended two illustrative examples—one labeled as misleading and the other as not misleading—before the test instance. These examples were designed to demonstrate the distinction between misleading and non-misleading thumbnails, offering practical context for how the model should reason through the task. Each example included a textual thumbnail description, a snippet of video subtitles, and a brief video description, along with the corresponding label and a concise explanation. By comparing these components, the model could better evaluate whether the test thumbnail aligned with the actual video content or relied on tactics such as exaggeration or clickbait. 
The following examples, one misleading and one not misleading, were included in the prompt to guide the model’s classification process.

\begin{quote}
\textbf{Example 1:} \\
\textbf{Thumbnail:} A person holding a stack of \$100 bills with the text “I made \$10,000 in one day!” \\
\textbf{Subtitles:} “In this video, I’ll share my experience of how I earned \$500 in a week through freelancing.” \\
\textbf{Video Description:} The creator discusses freelancing opportunities and shares how they earned \$500 in their first week. \\
\textbf{Categorization:} Misleading \\
\textbf{Explanation:} The thumbnail exaggerates the earnings (i.e., \$10,000 in a day) compared to the actual content (\$500 in a week), using clickbait tactics.

\vspace{1mm}

\textbf{Example 2:} \\
\textbf{Thumbnail:} A smiling chef holding a plate of pasta with the text “Easy 15-minute pasta recipe.” \\
\textbf{Subtitles:} “Today, we’re making a quick and delicious pasta dish that takes only 15 minutes to prepare.” \\
\textbf{Video Description:} The video demonstrates a step-by-step pasta recipe with a 15-minute prep time. \\
\textbf{Categorization:} Not Misleading \\
\textbf{Explanation:} The thumbnail accurately represents the content and preparation time.

\end{quote}

By providing these examples, we aimed to improve the models' ability to evaluate thumbnails consistently and accurately, helping them recognize subtle differences between misleading and non-misleading content. This structured approach gave the models a clear reference for making reliable assessments during the evaluation process.

\noindent \textbf{Dynamic Few-Shot Examples Prompt.} In the dynamic few-shot approach, we automatically selected two examples—one MTV and one NMTV—from the dataset that were semantically similar to the input video. Using a text-to-vector method, we analyzed and compared video descriptions to ensure that the chosen examples closely related to the video under evaluation.

\begin{itemize}[leftmargin=*,topsep=0pt] \setlength{\itemsep}{0pt} \setlength{\parskip}{0pt} 
\item \noindent \textbf{Text-to-Vector Conversion and Similarity Analysis:}
To compare videos based on semantic content, we used Sentence-BERT (SBERT), a transformer-based model that generates meaningful sentence embeddings via a Siamese architecture \cite{reimers2019sentencebertsentenceembeddingsusing}. Unlike standard BERT or RoBERTa, SBERT is highly efficient—reducing similarity search time among 10,000 sentences from ~65 hours to just 5 seconds—while preserving semantic accuracy. This makes it well-suited for retrieval tasks like our dynamic few-shot prompting setup.

To ensure fairness and eliminate bias in prompt construction, all video descriptions were generated uniformly using Twelve Labs, which was not part of the models being used for classification purposes. This consistency ensured that all models received equivalent semantic cues during dynamic few-shot prompting. We measured the semantic closeness of video descriptions using cosine similarity, which is more suitable than lexical metrics like BLEU or ROUGE for identifying paraphrased or semantically equivalent content. This enabled retrieval of videos that were contextually similar, not just lexically close. Using the computed similarity scores, we selected one video each from the misleading and non-misleading categories that most closely matched the input video, enabling a balanced and contextually relevant comparison set for downstream classification.
\item \noindent \textbf{Generating Thumbnail Descriptions and Explanations for Input Examples:}

We precomputed the thumbnail descriptions for all thumbnails in our dataset. Claude was selected to generate concise, one-sentence descriptions for each thumbnail due to its high accuracy in similar tasks. Since the example thumbnail images themselves were not included in the prompt, these descriptions served as the textual representation.

Next, we generated explanations for why a video’s thumbnail was categorized as misleading or not using Claude. With the thumbnail descriptions, ground truth labels, and truncated video descriptions and subtitles (each limited to 200 words), the model produced concise rationales for each classification. The 200-word cap was chosen to balance context and performance—empirical testing showed that longer inputs reduced accuracy, and prior work \cite{li2024longcontextllmsstrugglelong} suggests LLMs struggle with excessively long contexts. This limit also improved efficiency and aligned with the length of typical descriptions in our dataset.

\item \noindent \textbf{Incorporation of Examples into the Prompt:}

Each example followed a standardized format, featuring a brief textual description of the video’s thumbnail, truncated versions of both the video’s subtitles and video description (limited to 200 words each), and a categorization label specifying whether the thumbnail was ``Misleading" or ``Not Misleading," accompanied by an explanation. These examples provided clear and relevant reference points, helping the model evaluate and categorize thumbnails more accurately and consistently.

As discussed above, the 200-word limit was chosen to balance context and performance—those same considerations applied here for subtitles and video descriptions. Informed by empirical testing and prior work, this threshold helped preserve classification accuracy by avoiding the performance issues associated with excessively long inputs.

\end{itemize}
By implementing this dynamic few-shot strategy, we enhanced the model’s ability to assess whether a thumbnail was misleading or not. The inclusion of semantically similar and well-structured examples provided valuable context, enabling the model to make more accurate and consistent classifications. This approach contributed to the reliability of our findings and offered a more nuanced perspective on thumbnail evaluation.

\subsection{Models and Pipelines}

We employed four distinct pipelines for our evaluations, ensuring consistency across all platforms by using the same subtitles, thumbnails, and other supporting inputs. The only variation was in the video descriptions, which were generated by the respective LLM models. Specifically, \texttt{claude-3-5-sonnet@20240620} and \texttt{gemini-1.5-flash-001} generated their own video descriptions, while \texttt{gpt-4o-mini} and \texttt{gpt-4o} \cite{openai2024models} relied on video descriptions generated by Twelve Labs, as these models do not support direct video input. For \texttt{gpt-4o}, we used the default version, which can now be accessed via the model parameter: \texttt{gpt-4o-2024-05-13}, following the latest update. All models were evaluated using default temperature settings.

This approach provided a comprehensive evaluation of the models' performance, allowing us to determine which model best suited our use case. Additionally, we assessed both computational costs and accuracy to strike an optimal balance between performance and resource efficiency.

\section{Dataset Analysis}

This section presents an analysis of our curated dataset of 2,843 videos, comprising 1,359 MTVs and 1,484 NMTVs from eight countries. We examine category distributions, analyze the prevalence of various misleading tactics, and critically reflect on the dataset’s representativeness, including the measures taken to mitigate potential collection biases.

\subsection{Video Categories and Distribution}

\begin{figure}
    \centering
    \includegraphics[width=\linewidth]{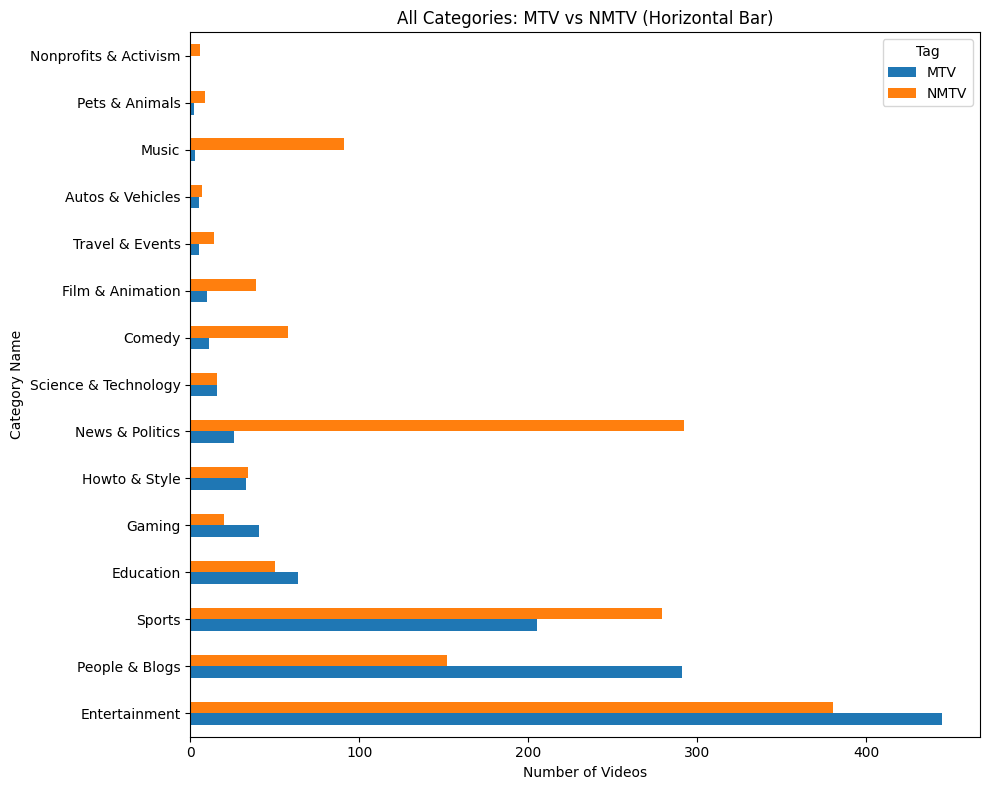}
    \caption{Overall categorical distribution of MTVs and NMTVs.}
    \label{fig:all_cat}
\end{figure}

Figure~\ref{fig:all_cat} presents the overall distribution of video categories in our dataset. The three most dominant categories—\textit{Entertainment}, \textit{Sports}, and \textit{People \& Blogs}—collectively account for the majority of both MTVs and NMTVs. This trend reflects broader YouTube engagement patterns, where these categories routinely attract high viewership and are commonly associated with misleading thumbnail practices. Their prominence underscores the importance of developing detection strategies that target high-volume content clusters.

While these categories dominate globally, we also observed notable regional variations. In developed regions \textit{Entertainment} is the dominant category for both MTVs and NMTVs, followed by \textit{People \& Blogs} and \textit{Sports}. The distribution shows a gradual decline across other categories, suggesting a stronger emphasis on personal narratives, lifestyle content, and culturally resonant media. In contrast, developing countries exhibit a more varied category distribution. While \textit{Entertainment}, \textit{Sports} and \textit{People \& Blogs} remain prevalent, we observe a significant increase in the \textit{News \& Politics} category, especially within NMTVs. This trend may reflect editorial norms or platform moderation practices that discourage misleading thumbnail use in news-related content. Additionally, categories such as \textit{Music}, \textit{Pets \& Animals}, and \textit{Science \& Technology} show moderate but balanced representation, indicating diverse content interests across regions.

These findings highlight the importance of region- and category-aware detection strategies that go beyond visual cues to incorporate cultural and contextual signals. To support this, we designed and evaluated our pipeline to enable both region-wise and category-wise analysis across the entire dataset. Tailoring models to such nuanced dimensions enhances robustness and improves generalizability in detecting misleading thumbnails across diverse global content landscapes.

\subsection{Dataset Bias and Selection Strategy}
We acknowledge that certain video types. particularly MTVs from entertainment-driven channels—are overrepresented in our dataset. However, this skew is not arbitrary; it reflects the real-world prevalence of misleading thumbnail tactics in high-traffic genres like entertainment and sports. Our aim was not to replicate YouTube’s global content distribution, but to capture misleading behavior where it naturally occurs at scale.

Since MTVs were identified through manual or “accidental” discovery—such as trending or random keyword searches, this reflects how misleading content typically surfaces on the platform. Rather than artificially flattening category distributions, we preserved these natural patterns, which align with documented engagement trends on YouTube.

Past studies have often relied on datasets heavily skewed toward NMTVs, limiting evaluation of false negatives.  Although our dataset does not fully represent YouTube’s overall video ecosystem, its balanced design enables rigorous testing across both MTVs and NMTVs, offering a reliable benchmark for evaluating misleading thumbnail detection.

\subsection{Approaches to Misleading Thumbnail Design}
Our analysis revealed various tactics used to create misleading thumbnails on YouTube. These tactics can be grouped into distinct categories:
\begin{itemize}[leftmargin=*,topsep=0pt]
\setlength{\itemsep}{0pt}
\setlength{\parskip}{0pt}
    \item \textbf{Exaggeration Tactics:} Thumbnails often exaggerate "before and after" scenarios, such as promoting rapid weight loss or anti-aging results.
    
    \item \textbf{Celebrity Manipulation:} Celebrities are frequently depicted in compromising situations, such as being in jail or a hospital, accompanied by fabricated dialogue bubbles portraying intense emotions.
    
    \item \textbf{Lifestyle Fantasies:} Many thumbnails showcase exaggerated luxurious lifestyles—cars, mansions, private planes—misleading viewers into believing the video content will mirror those images.
    
    \item \textbf{Fabricated Visuals:} Some thumbnails use manipulated images, such as merging human and animal features, or bold claims like "married" or "divorce confirmed" that are not substantiated by the video content.
    
    \item \textbf{Provocative and Sensational Language:} Words like "exclusive," curse words, and similar attention-grabbing terms are used, often without proper context.

    \item \textbf{Regional Trends:} In certain regions, a unique trend has emerged where users search for videos using only a period (full stop). These "full stop" videos often feature disturbing or creepy thumbnails, part of a meme-like search behavior on YouTube.
\end{itemize}

\subsection{Effectiveness of YouTube's Thumbnail Policy}
YouTube has a policy in place for handling misleading thumbnails, which can lead to their removal or, in more severe cases, the termination of an entire channel \cite{youtube2024thumbnails}. YouTube relies on user reports to flag these thumbnails, in addition to employing machine learning algorithms for detection \cite{youtube2024communityguidelines}. However, many misleading thumbnails go unreported by users, limiting the effectiveness of the current system. From our dataset of 1,359 MTV videos, the average video age was 442 days. Of these, the top 10 most viewed videos had an average age of 924 days, and only 65 videos were removed from the entire dataset over the course of seven weeks, highlighting the inefficiency of this approach in addressing the issue at scale.
\section{PERFORMANCE ANALYSIS OF LLMs}
We now turn to our evaluation of how LLMs perform in detecting misleading YouTube thumbnails. We compared model performance across different prompt types, analyzed the number of videos processed by each model, and evaluated the accuracy of their predictions. Additionally, we conducted a cross-country comparison to examine the performance of LLMs in detecting misleading thumbnails across different regions. To ensure the robustness of our findings, we benchmarked the results against existing standards. Lastly, we assessed the costs associated with each model to identify the most efficient and cost-effective solution for large-scale thumbnail analysis.

\subsection{Variation in Number of Processed Videos}

The number of misleading thumbnail videos processed varied across models due to differences in their filtering mechanisms. Google’s Gemini 1.5 Flash applied strict safety filters, which blocked potentially harmful content, leading to fewer videos being processed, especially in the MTV dataset \cite{google2024geminisafety}. These filters flagged content based on three primary enum codes: PROHIBITED\_CONTENT, which blocks material deemed too sensitive; SAFETY, which flags content related to issues such as hate speech or harassment; and RECITATION, which prevents unauthorized citations from being included in responses \cite{google2024safetyfilters}.

In contrast, Claude 3.5 Sonnet, GPT-4o, and GPT-4o Mini applied less restrictive filters, processing a greater number of videos but with a higher risk of allowing harmful content \cite{openai2024safety,anthropic2024aup}. Twelve Labs, while not using safety filters, excluded videos with resolutions below 360p, limiting its ability to process lower-quality content in certain regions. Ongoing research continues to enhance the reliability and effectiveness of these safety mechanisms \cite{kumar2024ethics,li2024safety}. For more details on the number of videos processed by each model, please refer to the Table~\ref{tab:videos_processed} in the Appendix.

\subsection{Comparison Across Models}

We conduct a comparative evaluation of four the models—Claude 3.5 Sonnet, Gemini 1.5 Flash, GPT-4o-mini, and GPT-4o—across key performance metrics, including accuracy, precision, recall, and specificity. This analysis highlights notable differences in their effectiveness, enabling a nuanced assessment of each model’s strengths and limitations.
\begin{figure} \centering 
\vspace{-0.3in}
\includegraphics[width=\linewidth]{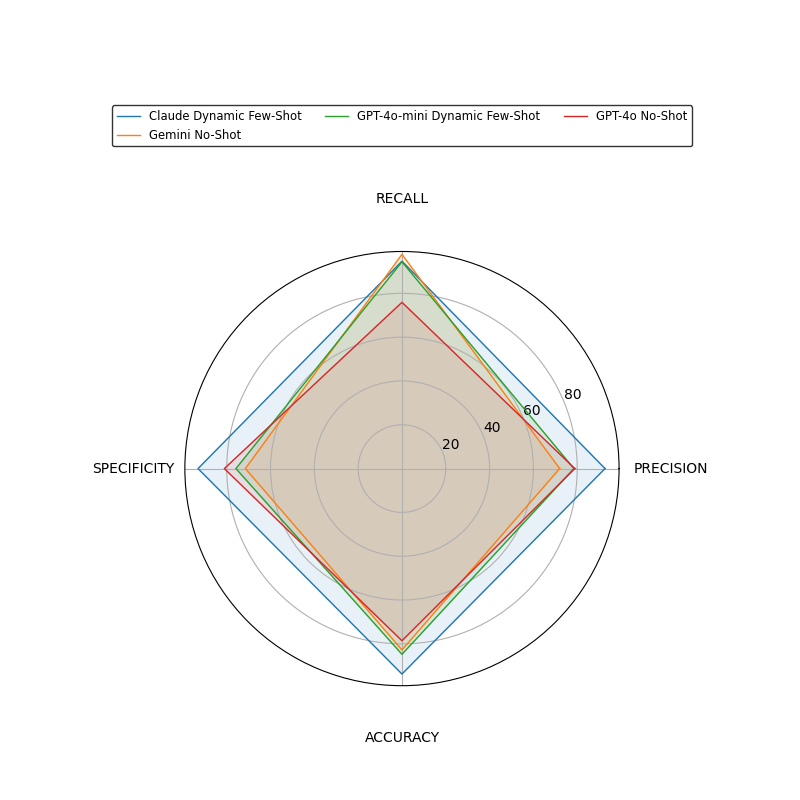} 
\vspace{-0.4in}
\caption{Radar plot of best prompt accuracy for four models across Accuracy, Recall, Precision, and Specificity.} 
\label{fig:radar_all} 
\end{figure}

\begin{table}[h!]
\centering
\begin{tabular}{|l|c|}
\hline
\textbf{Model}          & \textbf{Accuracy (\%)} \\ \hline
Claude 3.5 Sonnet       & 93.8             \\ \hline
Gemini 1.5 Flash        & 82.8             \\ \hline
GPT-4o Mini             & 84.8             \\ \hline
GPT-4o                  & 78.6            \\ \hline
\end{tabular}
\caption{Best accuracy of models across the three prompting strategies for the entire dataset.}
\label{tab:model_accuracy}
\end{table}
\noindent \textbf{Performance Overview.} 
As shown in Table~\ref{tab:model_accuracy}, the models exhibit clear differences in accuracy, highlighting variability in their ability to detect misleading thumbnails.

Claude 3.5 Sonnet consistently outperformed other models, achieving an accuracy of 93.8\% across the evaluation dataset, as shown in Figure~\ref{fig:radar_all}. A key contributor to Claude's superior performance is its robust chain-of-thought (CoT) reasoning capability. According to Anthropic’s official documentation \cite{anthropic2024chainofthought}, Claude is designed to naturally adopt structured, step-by-step reasoning—particularly when given open-ended or multi-component prompts.

In our setting, Claude—like all other models—was prompted to assess whether a thumbnail accurately represented the corresponding video, supported by subtitles and a generated video description. Even without explicit CoT instructions, Claude consistently produced well-organized outputs, frequently identifying visual-textual mismatches and synthesizing these into coherent justifications before issuing a classification.

For example, when evaluating a thumbnail featuring fantastical imagery (e.g., a ``banana cobra'' and a ``Spider-Man snail''), Claude generated the following explanation:

\begin{quote}
Based on the provided information, I would categorize this thumbnail as Misleading. Here's my explanation:

\textbf{Thumbnail content:} The thumbnail shows two unrelated images—a cobra snake made from a banana, and a snail-like creature with Spider-Man's appearance.

\textbf{Video content:} The video description and subtitles indicate that this is a compilation of various interesting and unusual scenes, including sunglasses, exercise equipment mishaps, recycling facilities, people walking on hot coals, and many other diverse topics. None of these match the images in the thumbnail.

\textbf{Lack of representation:} The thumbnail images are not mentioned anywhere in the video description or subtitles. There's no indication that the video actually includes content related to banana snakes or Spider-Man snails.

\textbf{Misleading expectations:} Viewers expecting to see content related to the thumbnail images would likely be disappointed, as the video seems to cover a wide range of unrelated topics.

The thumbnail is clearly designed to grab attention with its unusual and striking images, but it does not accurately represent the diverse compilation of scenes and activities actually present in the video. This discrepancy between the thumbnail and the actual content makes it misleading to potential viewers.

\end{quote}

Such responses exemplify Claude’s emergent CoT behavior, where reasoning is transparently laid out through clearly segmented observations. This capability significantly improves the model’s precision in detecting subtle mismatches and enhances its performance in borderline cases—such as distinguishing between similarly themed videos like interviews and dramatized clips—where contextual nuance is key.

Gemini 1.5 Flash excelled in recall, as shown in Figure \ref{fig:radar_all}, surpassing other models in detecting a wider range of misleading thumbnails. However, its performance in precision and specificity was weaker, particularly in cases requiring deeper contextual understanding. At times, Gemini's video descriptions missed critical details, failing to capture the full essence of the video content. In one example, the model misinterpreted a thumbnail featuring Jada Pinkett Smith, describing non-existent text as ``got go we." While it recognized the same celebrity in another case, it failed to accurately understand the thumbnail’s context. This inconsistency extended to other instances, such as when Gemini misidentified a thumbnail of Angelina Jolie from a movie scene and mistakenly labeled it as misleading. These examples highlight Gemini's challenges with celebrity recognition and its inability to align video descriptions with the actual content, leading to inaccurate conclusions.

GPT-4o-mini performed well, achieving an accuracy of 84.8\%. Although it trailed behind Claude, it outperformed Gemini in handling complex video content, showing a better grasp of subtle visual elements and intricate scenarios. However, it occasionally misclassified due to incorrect information from video descriptions, such as when Twelve Labs incorrectly identified "Prince Philip" instead of King Charles. Additionally, it failed to detect that a romantic scene in the thumbnail was fake and edited, missing its absence in the actual video. While it encountered fewer misclassifications compared to Gemini, GPT-4o-mini still struggled with distinguishing between real and fabricated thumbnails.

GPT-4o displayed mixed results, achieving an accuracy of 78.6\%, the lowest among the models. It struggled with interpreting dynamic visual cues and recognizing prominent public figures, which negatively impacted its overall performance. For example, GPT-4o failed to identify Lionel Messi in a misleading thumbnail of him arguing with a female referee, a scenario not present in the video itself. Additionally, the model misjudged a video based on a minor error in its description, resulting in an incorrect classification instead of a more comprehensive analysis. In another case, GPT-4o misclassified a thumbnail featuring an exaggerated image of an eagle carved from a watermelon, interpreting it as legitimate content rather than clickbait. These instances highlight GPT-4o's over-reliance on surface-level details, leading to errors, particularly with exaggerated or fabricated thumbnails.

Beyond accuracy, metrics like precision, recall, and specificity reveal important trade-offs in model behavior. Claude 3.5 Sonnet shows the highest specificity (0.931), effectively avoiding false positives—crucial for maintaining trust in content moderation. In contrast, Gemini 1.5 Flash achieves very high recall (0.978) but low specificity (0.715), often over-flagging non-misleading content. While acceptable in safety-critical contexts, this trade-off is less ideal for nuanced moderation. GPT-4o-mini with its moderate but balanced performance, whereas GPT-4o lags across all metrics. These results highlight the importance of specificity in real-world deployment, where minimizing false positives is essential for user trust and platform credibility.

\noindent \textbf{Key Differentiators.} Claude stood out not only for its superior accuracy but also for its ability to handle misleading thumbnails that were emotionally charged or visually exaggerated. While not perfect—especially when subtle manipulations were involved—it consistently outperformed the other models by detecting discrepancies between thumbnails and their corresponding video content. Gemini, despite excelling in identifying a wide range of misleading content, struggled with cases requiring deeper contextual understanding, particularly when public figures played a central role in the video narrative. GPT-4o-mini demonstrated strengths in handling nuanced prompts, surpassing GPT-4o, which consistently lagged behind. GPT-4o-mini performed effectively in scenarios where minor contextual variations were key to accurate classification, a task that Claude also excelled in but with even greater consistency. In contrast, GPT-4o lacked sensitivity to these variations, frequently missing critical elements needed for precise categorization.

\noindent \textbf{Limitations and Areas for Improvement.} All models had significant limitations in celebrity recognition. While Claude partially compensated with logical reasoning to infer content without explicitly identifying celebrities, other models struggled to recognize prominent figures, which led to misclassifications in scenarios where identification was essential. This limitation highlights a shared weakness across all models, except for Claude’s partial compensation through inference.

Claude could further improve by incorporating actual celebrity recognition to eliminate the need for inference, while Gemini would benefit from enhancing its ability to align visual content with broader video contexts. GPT-4o-mini, despite its overall robustness, could benefit from better detection of manipulated content. Finally, GPT-4o requires more significant improvements across the board, particularly in its ability to process dynamic and complex visual elements.

The overall performance comparison reveals that Claude 3.5 Sonnet sets a high benchmark in terms of accuracy, logical reasoning, and handling misleading thumbnails. GPT-4o-mini also shows potential, particularly in more complex video analysis scenarios. Gemini 1.5 Flash, while strong in recall, needs to improve its precision, especially when interpreting nuanced content. GPT-4o, though adequate in simpler tasks, lags behind in more challenging contexts, highlighting the need for further refinement in recognizing visual cues and public figures.

The integration of celebrity recognition and better handling of misleading visual cues could significantly improve the performance of all models, particularly Gemini and GPT-4o. As the field continues to evolve, fine-tuning these models to handle increasingly complex and dynamic thumbnails will be essential to enhance their effectiveness.

\subsection{Comparison Across Prompts}

\begin{figure}
    \centering
    \includegraphics[width=\linewidth]{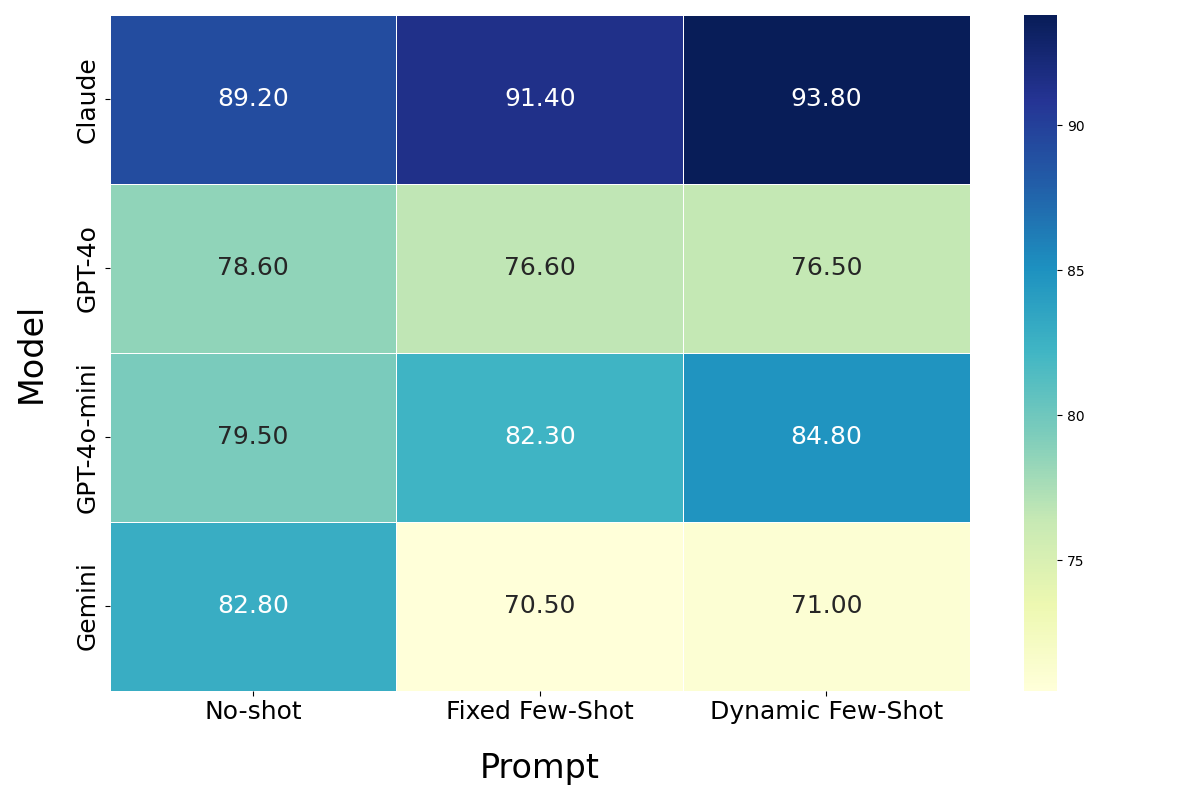}
    \caption{Model accuracies for each prompt}
    \label{fig:heat_map}
\end{figure}

We observed notable differences in accuracy when using different prompting strategies within the same model. Our hypothesis was that transitioning from \textit{no-shot} to \textit{fixed few-shot} and, finally, to \textit{dynamic few-shot} would lead to increased accuracy and improved overall metrics, as prior research suggested. This trend was evident in the top two models, Claude 3.5 Sonnet and GPT-4o-mini, which displayed consistently higher accuracies across prompts as observed in Figure \ref{fig:heat_map}.

For Claude 3.5 Sonnet, fluctuations in accuracy were observed with \textit{no-shot} prompting, with some regions falling below 90\% (see Appendix Figure \ref{fig1}). However, accuracy improved significantly with the introduction of \textit{few-shot} and \textit{dynamic prompting}. These improvements demonstrate the value of advanced prompting strategies, particularly when working with complex video thumbnails. By using these refined techniques, Claude's accuracy consistently exceeded 90\% across all regions.

In contrast, GPT-4o's performance remained relatively static across different prompting strategies, showing little improvement when moving from \textit{no-shot} to \textit{few-shot} prompting. Meanwhile, Gemini 1.5 Flash displayed a notable decrease in accuracy when moving from \textit{no-shot} to \textit{few-shot} prompts, as shown in Figure \ref{fig:heat_map}. This result aligned with our preliminary tests, which indicated that Gemini, while generally less effective than other models, performed relatively better when given simpler instructions, such as determining whether a video was misleading or not. It struggled when asked to handle more complex, step-by-step criteria for identifying misleading thumbnails. Although the performance declined with more detailed prompts, these findings provided valuable insights into the model’s behavior and highlighted areas for improvement in future applications.

\subsection{Performance Across Categories}

To evaluate how well the detection pipeline generalizes across diverse video categories, we analyzed the performance of Claude 3.5 Sonnet (with dynamic prompting) on a balanced subset of categories. We addressed class imbalance by sampling an equal number of MTVs and NMTVs for each category, using a 1:1 ratio based on the smaller class size (i.e., \texttt{min(total\_MTV, total\_NMTV)}), thereby ensuring fairness while preserving category diversity. Categories with no MTVs, such as \textit{Pets \& Animals} and \textit{Nonprofits \& Activism}, and those with very limited data ($\le 10$ videos after balancing, e.g., \textit{Autos \& Vehicles}) were excluded to maintain metric reliability and avoid misleading conclusions. Although most of these categories showed promising results, their low support made the metrics unreliable.


\begin{table}[h]
\centering
\scriptsize
\caption{Per-Category Accuracy and F1 Scores (Balanced Dataset)}
\label{tab:num_category_scores}
\begin{tabular}{|l|c|c|c|}
\hline
\textbf{Category Name} & \textbf{Accuracy} & \textbf{F1 Score} \\
\hline
Sports & 0.9530 & 0.9535 \\
Gaming & 0.9474 & 0.9500 \\
Education & 0.9388 & 0.9412 \\
Entertainment & 0.9107 & 0.9108 \\
Comedy & 0.9091 & 0.9091 \\
Howto \& Style & 0.9062 & 0.9032 \\
News \& Politics & 0.9038 & 0.8936 \\
Film \& Animation & 0.9000 & 0.8889 \\
People \& Blogs & 0.8826 & 0.8875 \\
Science \& Technology & 0.8750 & 0.8667 \\
\hline
\end{tabular}
\end{table}
As shown in Table~\ref{tab:num_category_scores}, the model demonstrates strong and consistent performance across all included categories, both in accuracy and F1 score. These results highlight the pipeline’s ability to generalize across varied, visually rich domains, and reinforce the importance of maintaining balanced and sufficiently sized category representations in future evaluations.

\subsection{Analysis of Misleading Thumbnails Across Countries}

\begin{figure} 
\centering 
\includegraphics[width=\linewidth]{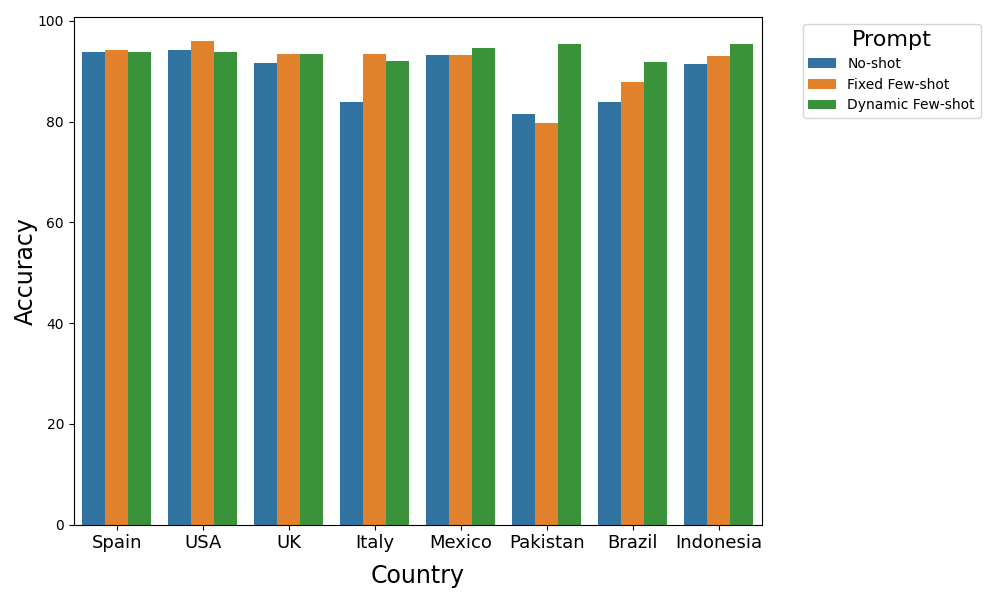} 
\vspace{-0.3in}
\caption{Accuracy using Claude 3.5 Sonnet for each prompt across all countries.} 
\label{fig:bestaccuracy_overall} 
\end{figure}

The models demonstrated varying effectiveness in detecting misleading thumbnails across different countries. On average, the detection accuracy for MTVs was similar between developed and developing countries, with accuracies of 82.3\% and 80.6\%, respectively. However, significant variations were observed with Claude's no-shot prompt, with some countries achieving over 93\% accuracy, while others, such as Italy, Brazil, and Pakistan, remained in the lower range, around 80\%. These misclassifications were primarily due to thumbnails featuring exaggerated content and sensational imagery. In Pakistan, there was an additional category of MTVs, particularly those centered around local celebrities in exaggerated scenarios intended to attract clicks. For example, in South Asia, many MTVs focused on film industry stars or cricket figures, reflecting a more localized style of clickbait.

These results suggest that beyond generic strategies for crafting such misleading thumbnails, regional differences in thumbnail design, cultural expectations, and language significantly influence the model's performance. The varied cultural and contextual cues embedded in thumbnails from different countries posed a challenge for the model to generalize across all regions.

To address these cultural nuances, dynamic few-shot prompting was employed. By incorporating culturally relevant examples into the prompt, as shown in Figure \ref{fig:bestaccuracy_overall}, accuracy improved significantly. In countries like Italy, Brazil, and Pakistan, where initial accuracy was relatively low, the use of prompts tailored to local contexts led to an increase in accuracy by at least 8\%, with all surpassing the 90\% mark. This demonstrates the importance of adapting models to regional contexts and providing more targeted input for better performance across diverse cultures.

\noindent \textbf{Common Strategies.} Despite regional differences, certain tactics were consistent across countries: \begin{itemize}[leftmargin=*,topsep=0pt] \setlength{\itemsep}{0pt} \setlength{\parskip}{0pt} \item \noindent Celebrities were universally leveraged—Pakistan and Spain often focused on sports stars, while the USA featured YouTubers and Hollywood actors. \item \noindent Thumbnails frequently exaggerated depictions of wealth or personal transformation, appealing to viewers' aspirations. \item \noindent Provocative or unrelated imagery was commonly used across all regions to drive clicks. \end{itemize}

\noindent \textbf{Key Differences.} Notable differences arose in how these strategies were executed across countries: \begin{itemize}[leftmargin=*,topsep=0pt] \setlength{\itemsep}{0pt} \setlength{\parskip}{0pt} \item \textit{Sensationalism and Clickbait:} Developing countries tended to use more bizarre, fantastical clickbait (e.g., absurd scenarios or fake hacks), whereas developed countries leaned towards more subtle sensationalism, such as exaggerated celebrity news or personal drama.

\item \textit{Sexualized Content:} Developing regions more frequently employed overtly provocative imagery to attract curiosity, particularly in societies where such content may be restricted. In contrast, developed countries tended to blend sexual content with satire or drama.

\item \textit{Wealth and Status:} In countries like Pakistan, exaggerated depictions of wealth and success were more common. Meanwhile, in the USA, emotional turmoil and public celebrity disputes often overshadowed wealth portrayal. \end{itemize}

In conclusion, addressing the varying strategies used for misleading content across different regions is key to improving model accuracy. Recognizing these cultural differences and incorporating localized data can significantly enhance the detection of misleading thumbnails across diverse contexts.


\subsection{Cost Breakdown and Optimization}
This section analyzes the costs associated with generating explanations, thumbnail descriptions, video descriptions, and classifying thumbnails using various models. The average video length in our dataset was 11.73 minutes, but videos longer than 30 minutes were truncated to 29 minutes 55 seconds, bringing the final average down to 10.21 minutes. Costs were calculated based on all inputs and outputs, including subtitles, video descriptions, and final classification.
Due to its superior performance, Claude was primarily used for generating explanations and thumbnail descriptions, which were later employed for dynamic few-shot prompting across all models. On average, generating a thumbnail description cost \$0.003805, while generating an explanation cost \$0.001637. For the full process of generating video descriptions and classifying thumbnails, the average cost for all prompts was \$0.0419 for Claude, and \$0.0161 for Gemini. For Twelve Labs the cost for generating video descriptions under their Developer Plan was on average \$0.437, with the average costs across prompts for classifying thumbnails at \$0.00703 for GPT-4o-mini and \$0.0529 for GPT-4o. Twelve Labs' Enterprise plan offers reduced rates for large-scale projects \cite{twelvelabs2024pricing}.

The costs for these evaluations are expected to decrease further as LLM usage becomes more affordable. The default GPT-4o model has already seen significant reductions in its latest release (\texttt{gpt-4o-2024-08-06}), with a 50\% decrease in input token costs and a 33\% reduction in output token costs. Additionally, batch processing and caching mechanisms across platforms will further lower expenses \cite{openai2024batchguide,anthropic2024messagebatches,anthropic2024promptcaching,google2024contextcache,openai2024promptcaching,google2024batchpredictiongemini}, improving both latency and cost-efficiency for such projects.


\subsection{Performance Benchmark Analysis}
Our results align with existing works on model benchmarks, confirming Claude 3.5 Sonnet's superior performance in classification tasks. Claude consistently demonstrated the highest accuracy in our study, with an average of 91.5\% across multiple prompts, maintaining low error rates even when handling complex and misleading thumbnails \cite{arshad2024ageval,whitbeckevaluating}. 

Gemini 1.5 Flash, while highly effective in blocking harmful content, processed fewer videos due to its strict safety filters, resulting in an average accuracy of 74.8\%. GPT-4o and GPT-4o-mini followed with average accuracies of 77.3\% and 82.2\%, respectively, showing competence but needing further improvement in managing complex or provocative thumbnails. Prior studies also support GPT-4o-mini’s stronger performance over GPT-4o and Gemini in intent classification and knowledge-based reasoning tasks \cite{maheshwari2024efficacy,sinha2024guiding}. Although Gemini excels in localized tasks such as temporal reasoning and summarization, it struggles with more complex, global tasks requiring deeper context understanding, yet remains competitive in shorter, visual tasks \cite{ataallah2024infinibench}.

\section{Comparison with Existing Work – CHECKER}

A central goal of our evaluation was to assess whether a modern LLM—used in a zero-training, inference-only setting—could outperform specialized multimodal pipelines designed for misleading thumbnail detection. To this end, we compared our best-performing configuration, \textbf{Claude 3.5 Sonnet with dynamic few-shot prompting}, against \textbf{CHECKER} \cite{xie2021checker}, a state-of-the-art model built specifically for this task.

CHECKER is a supervised multimodal framework that fuses visual and textual features (thumbnail + title) using advanced pooling mechanisms such as \textit{Block}, \textit{Mutan}, and \textit{MFH}. It employs a co-teaching strategy to mitigate the impact of noisy labels, and achieves strong performance on its 197-video test set (64 clickbait, 133 non-clickbait). Among all variants, \textit{CHECKER + Block ($\tau = 0.30$)} yielded the highest F1 score of 0.7153. However, when the same fusion technique was evaluated without access to weak supervision signals (i.e., generated labels), its performance dropped to 0.6538.

In contrast, our LLM-based pipeline required no fine-tuning or supervision. Despite operating in a purely inference-driven setting, Claude 3.5 Sonnet surpassed CHECKER’s best result with an F1 score of \textbf{0.7227}. 

We also include comparisons with baseline models evaluated in the CHECKER paper. As shown in Table~\ref{tab:checker_comparison}, Claude outperformed several vision-language transformer models—\textbf{VisualBERT}, \textbf{LXMERT}, and \textbf{UNITER}—which are pretrained to align textual and visual inputs by jointly encoding image features and text through transformer-based architectures. These models are commonly used in tasks like Visual Question Answering (VQA) and image-captioning. However, their reliance on object-centric image encoders makes them less effective for abstract or stylized content common in YouTube thumbnails, leading to weaker generalization in this domain.

We also compared against a traditional \textbf{Logistic Regression} baseline, which concatenates pre-extracted visual and textual embeddings but lacks any learned multimodal interaction or end-to-end optimization. As expected, it underperforms relative to both CHECKER and vision-language transformers due to its shallow architecture and limited representational power.

These results demonstrate that prompt-driven LLMs can match or exceed the performance of fully supervised, domain-specific architectures—offering a more flexible, training-free alternative. The success of Claude’s pipeline reinforces the viability of using large language models for content moderation tasks, especially when guided by structured reasoning prompts and rich contextual inputs.

\begin{table}[h]
\centering
\caption{F1 Score Comparison with CHECKER and Other Baselines}
\label{tab:checker_comparison}
\begin{tabular}{|l|c|}
\hline
\textbf{Model / Setup} & \textbf{F1 Score} \\
\hline
\textbf{Claude 3.5 Sonnet + Dynamic Examples} & \textbf{0.7227} \\
CHECKER + Block ($\tau = 0.30$) & 0.7153 \\
CHECKER + Block & 0.6831 \\
VisualBERT & 0.6722 \\
LXMERT & 0.6640 \\
UNITER & 0.6554 \\
CHECKER + Block (w/o gen. labels) & 0.6538 \\
Logistic Regression (with gen. labels) & 0.5986 \\
Logistic Regression (w/o gen. labels) & 0.4912 \\
\hline
\end{tabular}
\end{table}

\section{Ablation Study}

To evaluate the individual contributions of different textual modalities in our LLM-based detection pipeline, we conducted an ablation study using Claude 3.5 in a zero-shot setting. The goal was to isolate the impact of subtitles and video descriptions on classification performance, while maintaining a consistent prompt structure.

We performed the ablation only in the zero-shot setting to avoid altering few-shot exemplars, which rely on both subtitles and descriptions. Modifying these would introduce confounding factors, undermining the validity of the comparison.

The following input configurations were evaluated:

\begin{itemize}
    \item \textbf{Claude Zero-Shot}: Thumbnail + Description + Subtitles (complete input)
    \item \textbf{ABL-NS}: Thumbnail + Description (No Subtitles)
    \item \textbf{ABL-ND}: Thumbnail + Subtitles (No Description)
    \item \textbf{ABL-NDS}: Thumbnail only (No Description, No Subtitles)
\end{itemize}

\begin{table}[h]
\centering
\scriptsize
\caption{Ablation Study Results Using Claude 3.5 (Zero-Shot)}
\label{tab:ablation_results}
\begin{tabular}{|c|c|c|c|c|}
\hline
\textbf{Metric} & \textbf{ABL-NDS} & \textbf{ABL-ND} & \textbf{ABL-NS} & \textbf{Claude-Zero Shot} \\
\hline
Accuracy    & 0.8780 & 0.9077 & 0.9076 & 0.8920 \\
Recall      & 0.8010 & 0.8987 & 0.8856 & 0.8430 \\
Precision   & 0.9348 & 0.9016 & 0.9079 & 0.9240 \\
Specificity & 0.9487 & 0.9155 & 0.9258 & 0.9360 \\
\hline
\end{tabular}
\end{table}

As shown in Table~\ref{tab:ablation_results}, the best overall performance is observed in the full-input setting, \textbf{Claude-Zero Shot}, where all modalities—thumbnail, subtitles, and description—are present. Interestingly, \textbf{ABL-NS} (no subtitles) also yields strong performance, particularly in terms of accuracy and recall, and without any failure-to-classify cases. This suggests that video descriptions alone provide enough structured context for effective reasoning in many cases.

Removing both textual modalities (\textbf{ABL-NDS}) resulted in the lowest accuracy and recall, as the model relied solely on visual input. While specificity was highest in this configuration, this likely reflects a conservative bias due to lack of supporting context. In some cases, especially within the MTV subset, the model refused to classify thumbnails, citing ethical discomfort (e.g., “I do not feel comfortable analyzing this type of sensationalized content…”). These refusals reduced classification coverage and were excluded from metric computation.

In \textbf{ABL-ND}, where only the description is removed and the model relies on subtitles, we observe strong performance across most metrics. However, we also encountered a number of cases where the model failed to return a classification. This occurred when subtitles alone did not provide enough information, likely because some videos had sparse, low-quality or no subtitles. These incomplete responses were excluded from the final metric computation.

Overall, the results show that subtitles and descriptions offer complementary benefits. Subtitles improve the detection of specific misleading claims, while descriptions provide thematic grounding. Depending on the application, one may prioritize the higher recall of \textbf{ABL-NS} or the precision and specificity of the full-input \textbf{Claude Zero-Shot} configuration.

\section{Real-World Applicability}
Our proposed pipeline, available on the anonymized GitHub repository, is designed for seamless integration into YouTube’s existing infrastructure. It operates as a lightweight layer prior to the video upload process. The pipeline gathers input data—including the thumbnail, subtitles, and video description—and forwards it to a LLM for classification to assess whether the content is potentially misleading.

The model performs binary classification, labeling thumbnails as either misleading or not-misleading. This output can be integrated into YouTube’s upload workflow as a pre-screening mechanism. Upon video upload, the system analyzes the provided inputs. If the content is classified as not-misleading, the upload proceeds uninterrupted. If identified as misleading, the system can either flag the video for human moderation or temporarily block the upload, prompting content revision. This approach offers a proactive alternative to YouTube’s current post-hoc moderation, enabling early-stage intervention. By detecting misleading content at the point of upload, the platform can reduce the spread of deceptive media before it reaches users.

While our current implementation targets YouTube, the pipeline’s modular design makes it readily adaptable to other video platforms like Dailymotion, TikTok, and Instagram Reels. Future work will explore cross-platform generalizability and extend analysis across more content categories and popularity tiers.

\section{DISCUSSION}
\label{sec:discussion}

The deployment of such a system by video platform providers could significantly enhance content moderation efforts. However, both challenges and opportunities would need to be carefully considered:

\begin{itemize}[leftmargin=*,topsep=0pt]
\setlength{\itemsep}{0pt}
\setlength{\parskip}{0pt} 
\item \textbf{False Positive Mitigation:} While our LLM-based approach demonstrates high precision, even a small fraction of false positives could impact legitimate content creators. To address this, platforms could implement a multi-stage review process where flagged thumbnails undergo human review before any action is taken.
\item \textbf{Transparency and Appeals:} Clear communication about the use of AI-assisted moderation and an efficient appeals process would be crucial to maintain user trust and provide recourse for incorrectly flagged content.

\item \textbf{Cultural and Linguistic Context:} As Mohan and Punathambekar \cite{mohan2019localizing} highlight YouTube’s struggle to balance global and local strategies in linguistically diverse regions, LLMs may face similar challenges in regions lacking sufficient linguistic or cultural data, potentially impacting the accuracy of thumbnail classification.

\item \textbf{AI-Generated Content:} As AI-generated content gains popularity on platforms like YouTube \cite{hussain2024exploring}, there is an increasing risk of AI being used to create misleading thumbnails. Our proposed solution provides a critical safeguard against this growing issue, ensuring that AI-generated content remains accountable and responsible. 
\item \textbf{Adaptive Systems:} Given the evolving nature of online content, deploying this system as part of a continuous learning pipeline would allow for ongoing refinement based on new data and emerging trends in misleading content. This could be done through using updated LLMs, together with in-context learning or fine tuning.

\item \textbf{Regulatory Compliance:} 
 As regulations like the EU’s Digital Services Act (DSA) demand greater transparency and accountability \cite{eu_digital_services_act}, LLM-assisted moderation can help platforms like YouTube meet these obligations by improving detection and removal of harmful content. Regulatory bodies can also use this approach to audit platform compliance and enforce policy standards.
\end{itemize}

\section{Related Work}
Prior work on detecting misleading content on platforms like YouTube has largely focused on videos and associated metadata such as tags and titles. UCNet \cite{palod2019misleading}, OVCP \cite{shang2019towards}, and Bajaj et al. \cite{bajaj2016disinformation} rely heavily on user engagement or metadata signals, limiting their use to post-hoc detection. These approaches do not address the visual-semantic alignment of thumbnails with content, a key focus of our work.

CHECKER \cite{xie2021checker} and BaitRadar \cite{gamage2021baitradar} move toward thumbnail-based analysis but either rely on weak heuristics or omit actual video content. Our comparison with CHECKER demonstrates that LLM-based pipelines outperform such approaches. Furthermore, limitations in dataset availability (as in the case of BaitRadar) and data quality (as in CHECKER, which relies on crowdsourced annotations) further constrain meaningful comparability.

Recent studies have explored LLMs for automated content analysis \cite{gilardi2023chatgpt, gonzalez2024benchmarking}, while moderation tools like PIXELMOD \cite{paudel2024pixelmod} emphasize visual content. These align with our use of LLMs and highlight a growing shift toward more semantic, context-aware moderation approaches.

\textbf{Our Contribution.}  
We introduce a large-scale, cross-country dataset with balanced annotations by trained evaluators and propose an LLM-based pipeline that evaluates the semantic alignment between thumbnails and video content. Our approach addresses limitations in both dataset design and detection methodology found in earlier studies.

\section{CONCLUSION}

This paper presented a comprehensive analysis of misleading video thumbnails on YouTube, leveraging a large dataset and advanced LLMs to improve existing detection methods. Our approach demonstrated higher accuracy compared to traditional techniques relying on metadata and user comments. The findings highlight the need for more robust, scalable, and context-aware solutions to mitigate misleading content on video platforms. We recommend that platforms like YouTube enhance their enforcement mechanisms and transparency to protect viewers from misleading thumbnails and improve content consumption experiences.






\bibliographystyle{IEEEtran}
\bibliography{main}
%


\appendix
\section{Appendix}
\subsection{Supplementary Data}

The following tables and figures provide supplementary data that support the main findings of our study. These include detailed steps used for classifying YouTube thumbnails, the number of videos processed by each model, accuracy metrics for each model across different prompting strategies
 shown using radar plots for better visualization.

\begin{table}[ht]
\centering
\begin{tabular}{|c|p{7.4cm}|}  
\hline
\textbf{Step} & \textbf{Instruction} \\ \hline
1 & Analyze the video's thumbnail carefully. \\ \hline
2 & Read through the video’s subtitles to understand its content. \\ \hline
3 & Review the video’s textual description to gather context. \\ \hline
4 & Compare the thumbnail with the actual content from the subtitles and description. \\ \hline
5 & Determine if the thumbnail aligns with the video’s main topic or if it uses misleading tactics (e.g., exaggeration, false promises, or clickbait). \\ \hline
6 & Categorize the thumbnail as either "Misleading" or "Not Misleading." \\ \hline
7 & Provide a brief explanation justifying your classification decision. \\ \hline
\end{tabular}
\caption{Steps for Classifying YouTube Thumbnails}
\label{tab:classification_steps}
\end{table}

Table \ref{tab:classification_steps} outlines the process followed for classifying YouTube thumbnails as either "Misleading" or "Not Misleading." Each step was carefully designed to ensure a comprehensive evaluation of the thumbnail in relation to the video's actual content.

\begin{table}[ht]
\centering
\begin{tabular}{|c|c|}
\hline
\textbf{Model}      & \textbf{Videos Processed} \\ \hline
Claude              & 2759                      \\ \hline
Gemini              & 2135                      \\ \hline
GPT-4o-mini \& Twelve Labs & 2769              \\ \hline
GPT-4o  \& Twelve Labs     & 2749              \\ \hline
\end{tabular}
\caption{Average Number of Videos Processed by Each Model}
\label{tab:videos_processed}
\end{table}

As seen in Table \ref{tab:videos_processed}, each model processed a varying number of videos. The difference in video processing capacity across models helped assess their differing performance for same inputs.


Table \ref{tab:model_accuracy} summarizes the best accuracy achieved by each model across the three prompting strategies. This data helps to compare the overall effectiveness of the models in detecting misleading thumbnails.

\begin{figure}[h] 
\centering 
\includegraphics[width=\linewidth]{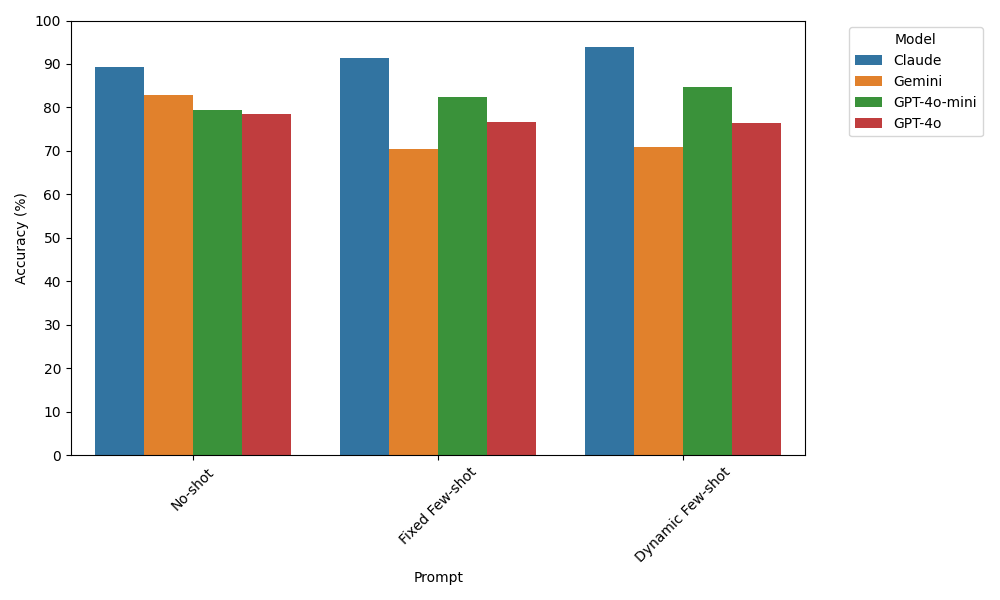} 
\caption{Graph comparing model performances across prompts} 
\label{fig:overall_accuracy} 
\end{figure}

Figure \ref{fig:overall_accuracy} provides a graphical comparison of model accuracies across different prompting strategies. This figure highlights the variations in model accuracy.

In addition to the accuracy metrics, the following radar plots offer a visual comparison of the models' performance across four key metrics: Accuracy, Recall, Precision, and Specificity.

\begin{figure}[h] 
\centering 
\includegraphics[width=\linewidth]{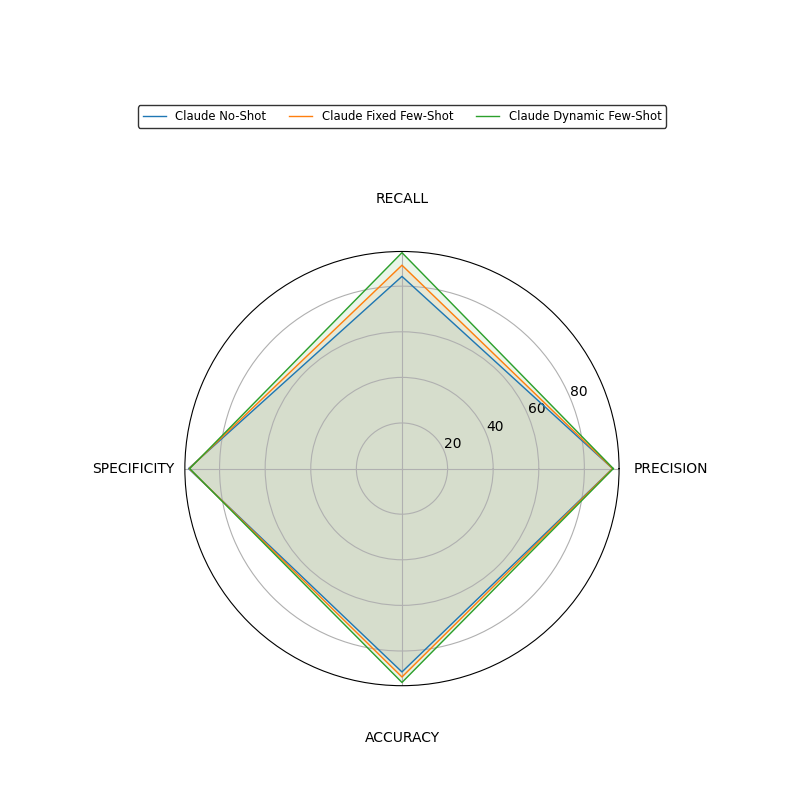} 
\caption{Radar plot for Claude for all prompts} 
\label{fig:radar_claude} 
\end{figure}

Figure \ref{fig:radar_claude} illustrates the radar plot for Claude, showcasing its performance across four metrics: Accuracy, Recall, Precision, and Specificity.

\begin{figure}[h] 
\centering 
\includegraphics[width=\linewidth]{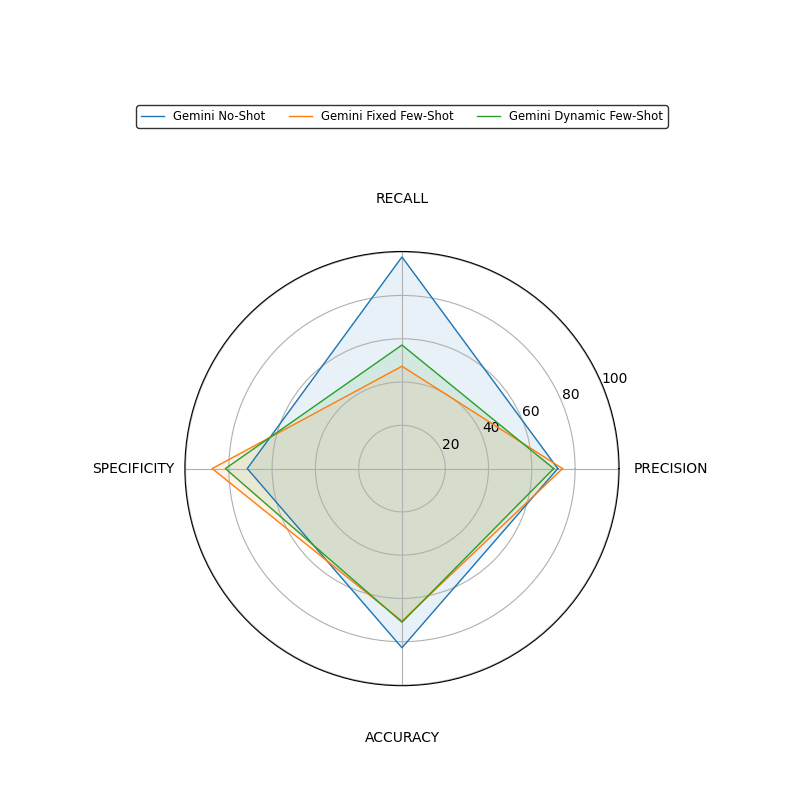} 
\caption{Radar plot for Gemini for all prompts} 
\label{fig:radar_gemini} 
\end{figure}

Similarly, Figure \ref{fig:radar_gemini} demonstrates Gemini's performance using the same metrics. This helps visualize the strengths and weaknesses of each model in handling misleading video thumbnails.

\begin{figure}[h] 
\centering 
\includegraphics[width=\linewidth]{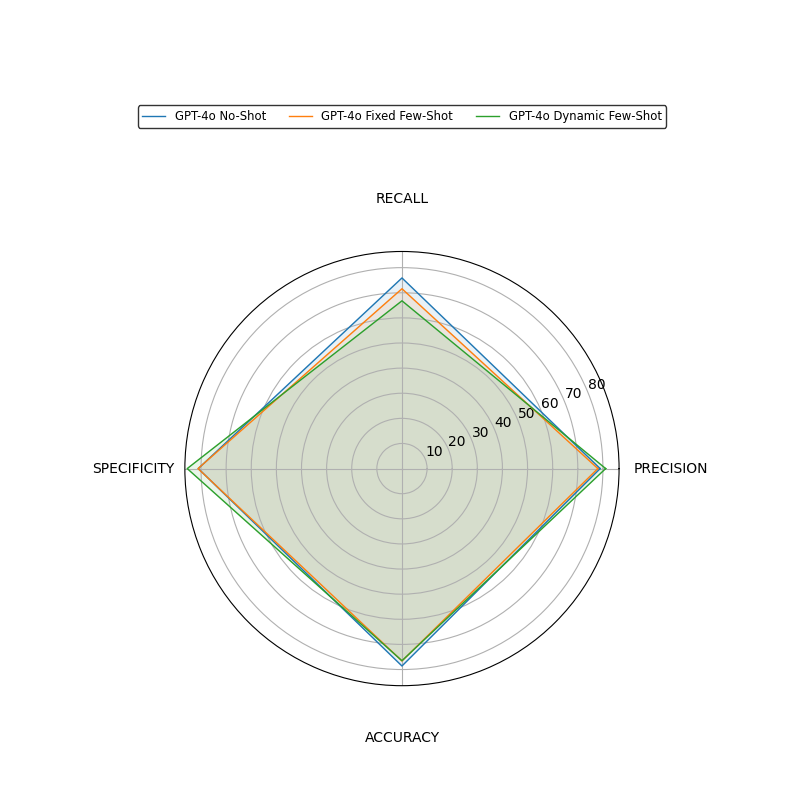} 
\caption{Radar plot for GPT-4o for all prompts} 
\label{fig:radar_gpt4o} 
\end{figure}

\begin{figure}[h] 
\centering 
\includegraphics[width=\linewidth]{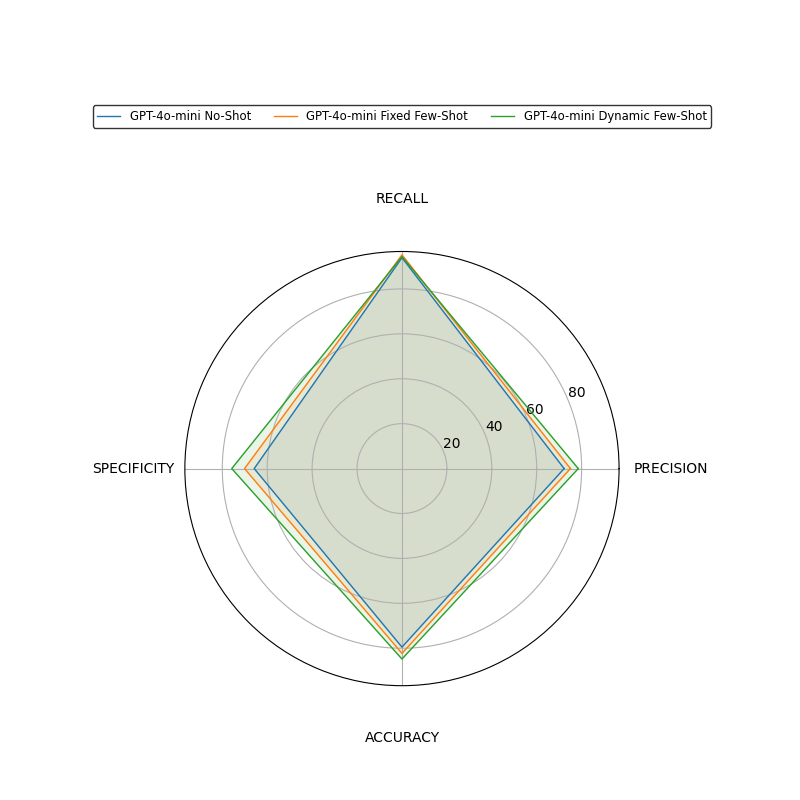} 
\caption{Radar plot for GPT-4o-mini for all prompts} 
\label{fig:radar_gpt4omini} 
\end{figure}

Figures \ref{fig:radar_gpt4o} and \ref{fig:radar_gpt4omini} represent the performance of GPT-4o and GPT-4o-mini, respectively. These figures allow for direct comparisons between the models based on the defined performance metrics.

\begin{figure}[h] 
\centering 
\includegraphics[width=\linewidth]{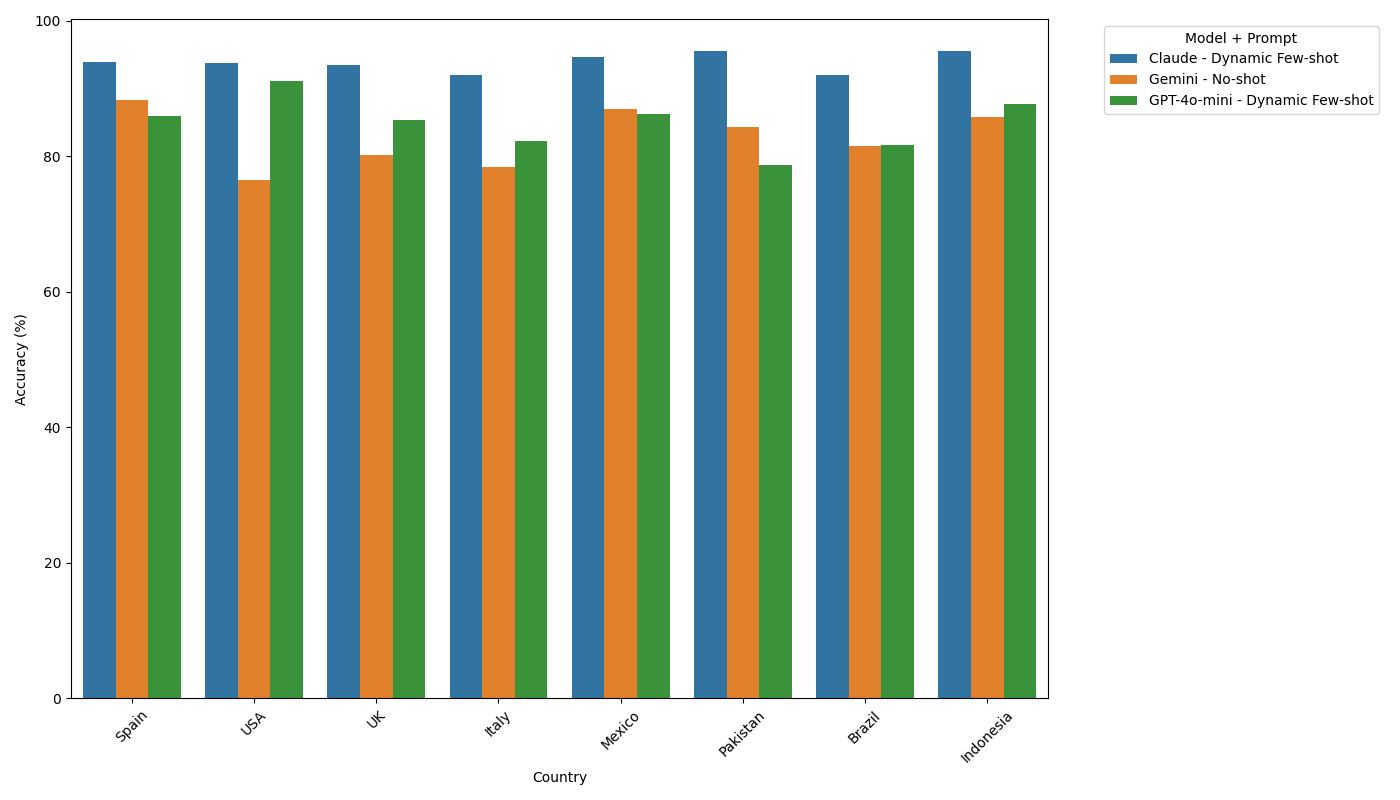} 
\caption{Top 3 Models with their Highest Overall Accuracy} 
\label{fig:bestaccuracy_top3models} 
\end{figure}

Figures \ref{fig:bestaccuracy_top3models} displays the top 3 models with the highest accuracy among the three prompts.

\begin{figure}[h] \centering \includegraphics[width=\linewidth]{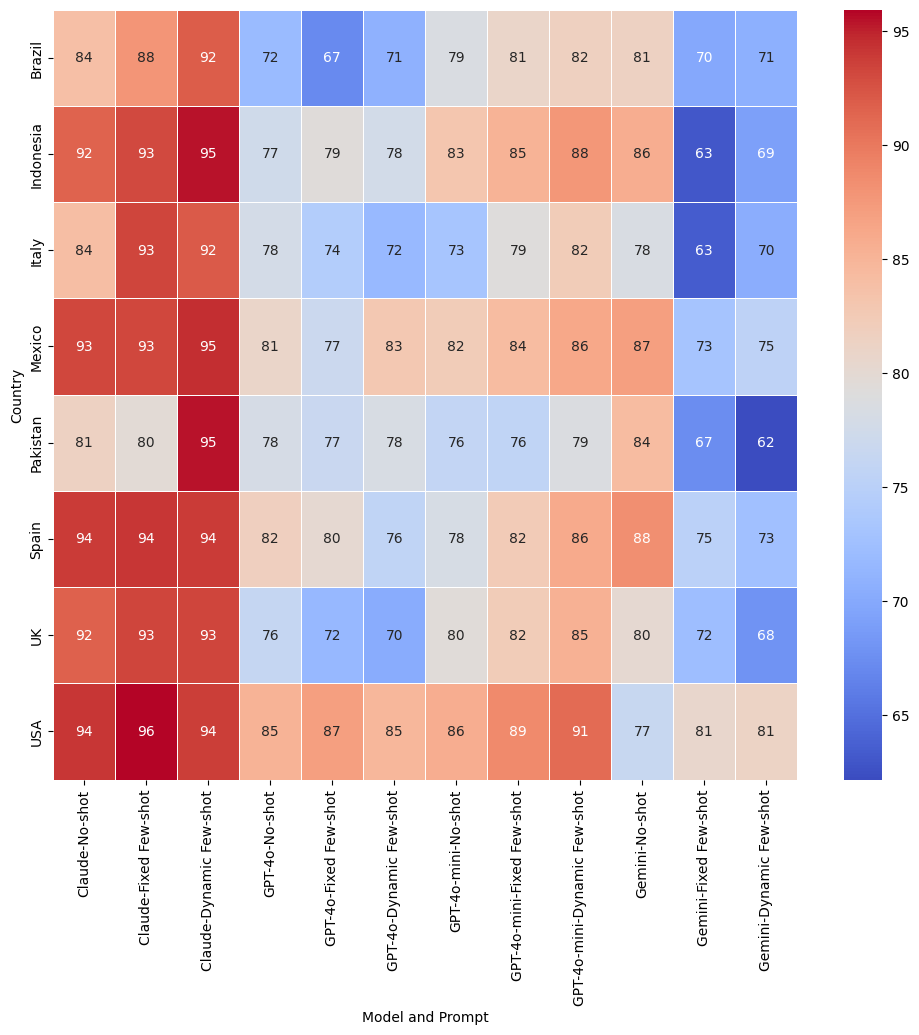} \caption{Heatmap for Model Accuracies Across Prompts for Each Country} \label{fig
} \end{figure}

\begin{figure}[h]
    \centering
    \includegraphics[width=\linewidth]{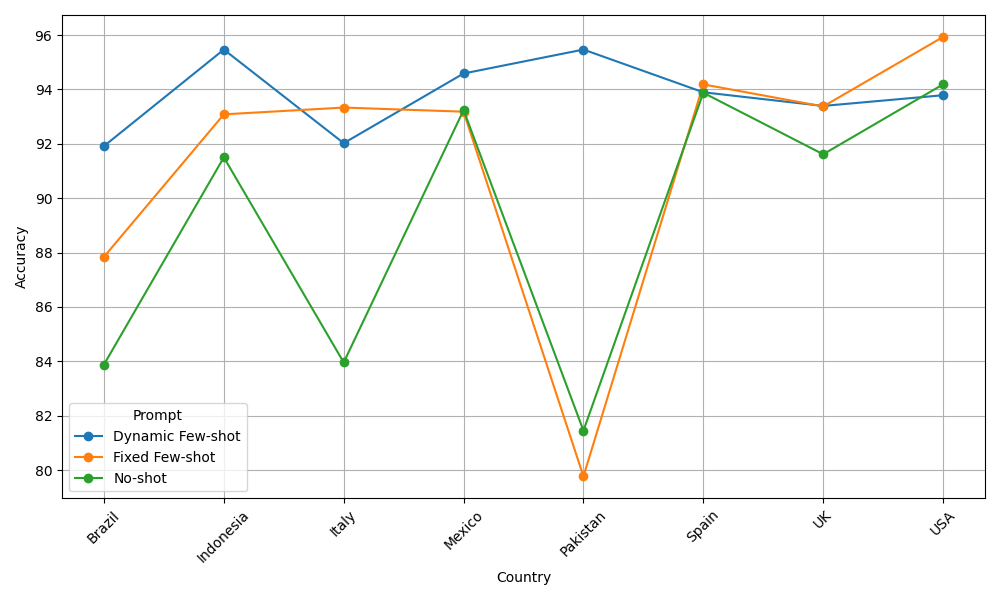}
    \caption{Claude Accuracies Across Prompts for Each Country}
    \label{fig1}
\end{figure}

\end{document}